\documentclass[12pt]{article}
\usepackage{amsmath}
\usepackage{amsfonts}
\usepackage{amssymb}
\usepackage{mathrsfs}
\usepackage{multicol}
\usepackage{color}
\usepackage{graphicx}
\usepackage{pstricks}
\usepackage{pst-node}
\usepackage{wrapfig}
\usepackage{enumerate}
\usepackage{caption}
\usepackage{cite}
\usepackage{subfigure}
\usepackage{amsfonts,amssymb}
\usepackage{amsthm}
\usepackage{amscd}
\definecolor{gray1}{gray}{0.1}
\definecolor{gray2}{gray}{0.2}
\definecolor{gray3}{gray}{0.3}
\definecolor{gray4}{gray}{0.4}
\definecolor{gray5}{gray}{0.5}
\definecolor{gray6}{gray}{0.6}
\definecolor{gray7}{gray}{0.7}
\definecolor{gray8}{gray}{0.8}
\definecolor{gray9}{gray}{0.9}

\newpsobject{showgrid}{psgrid}{subgriddiv=1,griddots=10,gridlabels=6pt}

\definecolor{dark-green}{rgb}{0,0.7,0}
\definecolor{dark-blue}{rgb}{0,0.2,0.5}
\definecolor{med-blue}{rgb}{0,0.7,1}
\definecolor{mblue}{rgb}{0,0.2,1}
\definecolor{cnc}{rgb}{0.8,0,0}
\definecolor{light-red}{rgb}{1,0.8,0.8}
\definecolor{dark-yelow}{rgb}{1,0.8,0}
\definecolor{light-blue}{rgb}{0.8,0.9,1}
\definecolor{verylight-blue}{rgb}{0.93,0.95,1}
\definecolor{light-yelow}{rgb}{1,0.9,0.8}
\definecolor{grey}{gray}{0.88}

\newpsobject{showgrid}{psgrid}{subgriddiv=1,griddots=10,gridlabels=6pt}

\begin{document}

\thispagestyle{empty}

\setlength{\abovecaptionskip}{10pt}

\begin{center}
{\Large\bfseries\sffamily{Spontaneous scalarization of boson stars}}
\end{center}
\vskip 3mm
\begin{center}
{\bfseries{\sffamily{Yves Brihaye$^{\rm 1}$, and Betti Hartmann$^{\rm 2}$ }}\\
\vskip 3mm{$^{\rm 1}$\normalsize Physique-Math\'ematique, Universit\'e de Mons-Hainaut, 7000 Mons, Belgium}\\
{$^{\rm 2}$\normalsize{Instituto de F\'isica de S\~ao Carlos (IFSC), Universidade de S\~ao Paulo (USP), CP 369, \\
13560-970 , S\~ao Carlos, SP, Brazil}}}
\end{center}

\begin{abstract} 
We study the spontaneous scalarization of spherically symmetric, asymptotically flat boson stars in the
$(\alpha {\cal R} + \gamma {\cal G}) \phi^2$ scalar-tensor gravity model. These compact objects
are made of a complex valued scalar field that has harmonic time dependence, while their space-time
is static and they can reach densities and masses similar to that of supermassive black holes.
We find that boson stars can be scalarized for both signs of the scalar-tensor coupling $\alpha$ and $\gamma$, respectively. This is, in particular, true for boson stars that are {\it a priori} stable with respect to decay into individual bosonic particles. A fundamental difference between the $\alpha$- and $\gamma$-scalarization exists, though:
while we find an interval in $\alpha > 0$ for which boson stars can {\it never} be scalarized when $\gamma=0$,
there is no restriction on $\gamma\neq 0$ when $\alpha=0$. 
Typically, two branches of solutions exist that differ in the way the boson star gets scalarized: either the
scalar field is maximal at the center of the star, or on a shell with finite
radius which roughly corresponds to the outer radius of the boson star. We also demonstrate that the 
former solutions can be radially excited. 
\end{abstract}

\section{Introduction}
``Compact objects''  is the name given to solutions in General Relativity (or alternative models of gravity)
that are typically so dense that the curvature of space-time around them has detectable effects. A ``measure'' of the  strength of the gravitational field of a body of mass $M$ and radius $R$ for $r > R$ can be given by comparing the dimensionless ratio $2GM/(c^2 R)$ to unity, where $G$ is Newton's constant
and $c$ the speed of light.  For a white dwarf \cite{ref1} and neutron star (for a recent review see e.g. \cite{ref2}), this ratio is on the order
of $2GM_{{\rm WD}}/(c^2 R_{{\rm WD}})\sim {\rm O}(10^{-8})$ and  
$2GM_{{\rm NS}}/(c^2 R_{{\rm NS}})\sim {\rm O}(10^{-1})$, respectively, but objects exist that are so dense, that $R\equiv r_h=c^2/(2GM)$. These objects are called {\it black holes} \cite{ref3}, because the surface associated to 
$r=r_h$ is a surface of infinite redshift for an asymptotic observer, called {\it event horizon}. This means
that any signal emitted from $r \leq r_h$ cannot reach an asymptotic observer. 

While neutron stars have gravitational fields comparable to that of black holes, limits on their upper
mass exist that are on the order of a few solar masses. Hence, they cannot account for the objects in
the centre of galaxies that possess millions of solar masses in a spatially small region. It seems that
the only viable and “standard” option are supermassive black holes \cite{ref4} and this seems to
have been confirmed by the recent observation of the center of M87 by the {\it Event Horizon telescope} \cite{ref4_new}. 
An alternative, albeit more exotic, are compact objects made out of bosonic fields, the simplest
example being the boson star made of a complex valued, massive scalar field \cite{ref6,ref7,ref8,ref9,ref10,ref11,ref12}. The global U(1)
symmetry in the model leads to a locally conserved Noether current and a globally conserved Noether
charge $Q$, where the latter can be interpreted as the number of scalar bosons of a given mass that
form the boson star in a self-gravitating, bound system. And, indeed, these objects have early on been
discussed as viable alternatives for the centres of galaxies \cite{ref13,ref14} and have - up to today - not been excluded 
(see also \cite{ref5} for a recent discussion). Since the discovery of a fundamental
scalar field in nature, the Higgs field \cite{ref15}, and the fact that scalar fields are extensively used in diverse
areas of physics, these objects can certainly be considered to have a probability to exist in nature or, at
least, to effectively describe very dense states of matter.
The study of compact objects as dense as black holes and boson stars has interest in its own right,
but is also very important from another perspective : since these objects have very strong gravitational
fields, they are also an ideal testing ground for alternative models of gravity and/or the limits of General
Relativity, which works extremely well in the weak regime, but has not been explored in full detail in
the regime of very strong gravity.
Because of the reasons mentioned above for the re-newed interest in scalar fields, scalar-tensor gravity
models are discussed extensively currently, in particular in the context of the primordial universe and its
exponential expansion very shortly after the Planck era, referred to as “inflation”. Horndeski scalar-tensor
gravity models \cite{ref16} lead – like General Relativity – to equations of motion that are maximally of second
order in derivatives \cite{ref17,ref18}. Black holes have been studied in models that possess a shift symmetry for the
scalar field \cite{ref19,ref20} as well as in models that allow spontaneous scalarization of black holes in the sense
that black holes carry scalar hair for sufficiently large values of the non-minimal coupling constants. In
general, the models contain terms of the form $f(\phi){\cal I}(g_{\mu\nu}; \Sigma)$, where $f(\phi)$ is a function of the scalar field
and ${\cal I}$ depends on the metric $g_{\mu\nu}$ and/or other fields $\Sigma$. 
Models have been studied with $f(\phi)\sim \phi^2$ \cite{ref21,ref22,ref23,ref24}
with different other forms of $f(\phi)$ with a single term in $f(\phi)$ \cite{ref25,ref26,ref27} as well as a combination of different
powers of $\phi$ \cite{ref28}. In all these cases, ${\cal I}$ has been chosen to be the Gauss-Bonnet (GB) term ${\cal G}$. 
In \cite{ref29} a model combining the original shift symmetric scalar field and a quadratic scalar field coupled to the GB
term has been studied bridging between shift symmetry and spontaneous scalarization. All the above
mentioned studies considered uncharged black holes, but charge can be included and leads to new effects.
The first studies of this type have been done in the conformally coupled scalar field case \cite{ref21} as well as
for the $f(\phi){\cal G}$ case \cite{ref30}. The studies were extended to discuss the approach to the extremal limit in \cite{ref31} and it was demonstrated that scalarized black holes can exist for
both signs of the scalar-tensor coupling due to the fact that ${\cal G}$ becomes negative close to
the horizon when approaching extremality.
Since the electromagnetic field can source the scalar field when non-minimally coupled, charged black
holes can also carry scalar hair without scalar-tensor coupling \cite{ref32,ref33}.

The first study of neutron stars in scalar-tensor gravities has been done in \cite{ref1new}. The energy-momentum tensor of the spherically symmetric, static neutron stars was assumed to be that of a perfect fluid. Spontaneous scalarization
of neutron stars appears when the function that couples the scalar and tensor part of gravity fulfils
an inequality. In particular, this means that for neutron stars with large mass it is energetically 
favourable to develop a non-constant value of the scalar field inside their core. 
A similar study was done for boson stars that possess
an energy-momentum tensor that (in general) is not of perfect fluid type \cite{ref2new}. While in this latter
paper, it was claimed that spontaneous scalarization does not exist without self-interaction,
a study of the dynamical process of scalarization of a boson star \cite{ref3new} demonstrated that
a mass term for the scalar field forming the boson star is sufficient for this phenomenon to appear.

In this paper, we study boson stars in scalar-tensor gravity with a coupling between the gravity scalar and the Ricci scalar
${\cal R}$ and/or the Gauss-Bonnet term ${\cal G}$, respectively. One of the coupling functions studied in \cite{ref2new,ref3new} is equal to the one we are studying here in the limit
of vanishing coupling to the Gauss-Bonnet term. We point out, however, that boson stars can be scalarized 
for both signs of the scalar-tensor coupling and that -- in addition -- radial excitations of the scalar field
exist.
In Section 2, we give the model, while in Section 3 we
discuss the tachyonic instability appearing in the system.
In Section 4, we discuss the full backreacted system, and Section 5 contains
are conclusions. In the following, we will use units such that $c\equiv 1$.

\section{The model}
The model we are studying here is a scalar-tensor gravity model that contains a non-minimal coupling between the square of a  real scalar field and the Ricci scalar ${\cal R}$ as well as the Gauss-Bonnet term ${\cal G}$~:
\begin{equation}
\label{action}
S =  \int  {\rm d}^4x  \sqrt{-g} \left[\frac{{\cal R}}{2} + (\alpha {\cal R} +\gamma {\cal G})\phi^2 -  
\frac{1}{2}\partial_{\mu} \phi  \partial^{\mu} \phi  + 8\pi G {\cal L}_{\Psi} \right]   \ ,
\end{equation}
where 
${\cal L}_{\Psi}$  corresponds to the Lagrangian density of a complex valued, massive scalar field~:
\begin{equation}
{\cal L}_{\Psi}=-\left(\partial_{\mu} \Psi\right)^* \partial^{\mu}\Psi - m^2\Psi^* \Psi \ ,
\end{equation}
and ${\cal G}$ is the Gauss-Bonnet term
\begin{equation}
{\cal G}=R_{\mu\nu\rho\sigma} R^{\mu\nu\rho\sigma} - 4 R_{\mu\nu} R^{\mu\nu} + R^2   \ .
\end{equation}
$m$ is the mass of the scalar field $\Psi$ and $G$ Newton's constant, respectively, and from now on we will choose 
units such that $8\pi G\equiv 1$. Variation of the action (\ref{action}) with respect to the metric, the complex
valued scalar field $\Psi$ and the real valued gravity scalar $\phi$ leads to the following set of coupled, non-linear
differential equations~:
\begin{eqnarray}
\label{eom}
G_{\mu\nu}=T^{(\Psi)}_{\mu\nu} + T^{(\phi)}_{\mu\nu} \ \ \ , \ \ \ (\square - m)\Psi=0 \ \ \ \ , \ \ \ 
\left(\square +\alpha {\cal R}+\gamma{\cal G}\right)\phi=0   \ ,
\end{eqnarray}
where
\begin{equation}
T^{(\Psi)}_{\mu\nu} = g_{\mu\nu} {\cal L}_{\Psi} - 2 \frac{\partial L_{\Psi}}{\partial g^{\mu\nu}}=
-g_{\mu\nu}\left[\frac{1}{2} g^{\alpha\beta} \left(\partial_{\alpha} \Psi^* \partial_{\beta}\Psi +
\partial_{\beta} \Psi^* \partial_{\alpha} \Psi\right) + m^2 \Psi^* \Psi\right] +
\partial_{\mu} \Psi^* \partial_{\nu} \Psi + \partial_{\nu} \Psi^* \partial_{\mu} \Psi
\end{equation}
is the energy-momentum tensor associated to the boson star field $\Psi$ and
\begin{eqnarray}
T^{(\phi)}_{\mu\nu}&=&\frac{1}{2}\partial_{\mu}\phi \partial_{\nu}\phi - \frac{1}{4} g_{\mu\nu} \partial_{\sigma}\phi \partial^{\sigma}\phi + 2\alpha \left[D_{\mu}(\phi \partial_{\nu} \phi) - g_{\mu\nu} D_{\sigma} (\phi \partial^{\sigma} \phi)\right] \nonumber \\
&-& \gamma \left(g_{\rho\mu} g_{\lambda\nu} + g_{\lambda\mu} g_{\rho\nu}\right)
\eta^{\kappa\lambda\beta\delta}\eta^{\rho\iota\sigma\tau} R_{\sigma\tau\beta\delta} D_{\iota} (\phi\partial_{\kappa}\phi)
\end{eqnarray}
the tensor associated to the $\phi$ field. 

Note that the action
possesses a global $U(1)$ symmetry of the form $\Psi \rightarrow \exp(i\chi)\Psi$, where $\chi\in \mathbb{R}$ is a constant, but arbitrary phase. This leads to the existence of a globally conserved Noether charge $Q$, which can be interpreted as the number of bosonic particles of mass $m$
that form a self-gravitating, bound system which is referred to as a {\it boson star}.

In contrast to spherically symmetric, electro-vacuum black holes, boson stars have been shown to scalarize
even when $\gamma=0$ \cite{ref2new,ref3new}. This is possible, because the energy-momentum tensor of the $\Psi$-field has (in general) a non-vanishing trace ${\cal T}^{(\Psi)}\equiv g^{\mu\nu} T^{(\Psi)}_{\mu\nu}$. In the perturbative limit
(i.e. when neglecting backreaction of the scalar $\phi$ on the space-time), the Einstein equation tells us that  
${\cal R}= -{\cal T}^{(\Psi)}$ and hence, we would expect a tachyonic instability to be possible through scalar-tensor coupling of the $\alpha\phi^2 {\cal R}$ type. Note that this is fundamentally different to the
case of spherically symmetric, asymptotically flat electro-vacuum black holes for which
${\cal R}=0$. In this latter case, it is thence necessary to couple the scalar field differently, e.g. to the Gauss-Bonnet term, in order to achieve spontaneous scalarization. 

In the following we will discuss spherically symmetric, non-rotating
boson stars, which are solutions to (\ref{eom}) for $\alpha=\gamma=0$.
For the metric and scalar fields, we choose the following Ansatz~:
\begin{equation}
\label{eq:metric}
ds^2 = - N(r)\sigma(r)^2 dt^2 + \frac{1}{N(r)} dr^2 + r^2 \left(d\theta^2 + \sin^2\theta d\varphi^2\right) \ \ ,  \ \ \phi=\phi(r)
\end{equation}
while the complex scalar field forming the boson star reads~:
\begin{equation}
\Psi(r,t)=\exp(i\omega t) \psi(r) \ .
\end{equation}
with $\omega$ a real and positive constant.

The boson stars will be characterized by their mass $M$ and Noether charge $Q$. The former can be read off from the behaviour
of the metric function $N(r)$ at infinity~:
\begin{equation}
\label{asymp_bs_mass}
     N(r \rightarrow \infty) \rightarrow 1 - \frac{2M}{8 \pi r}  + {\rm O}\left(\frac{1}{r^2}\right)  \ ,
\end{equation}
while the Noether charge in our coordinate system reads~:
\begin{equation}
\label{noether_charge}
        Q =  8 \pi \omega \int\limits_0^{\infty}  \frac{r^2 \psi^2}{N \sigma} \ {\rm d}r \ .
\end{equation}
When a non-trivial scalar field $\phi$ exists, there is an additional charge that characterizes the solutions,
the scalar charge $Q_{\phi}$, which appears in the power-law fall-off of the scalar field function~:
\begin{equation}
\phi(r\rightarrow\infty)\sim\frac{Q_{\phi}}{r} \ .
\end{equation}

In the following, we will first demonstrate that a tachyonic instability for the scalar field $\phi$ exists in
the background of the boson star space-time. After that we will solve the full back-reacted system in the limit
$\gamma=0$ and demonstrate that fully backreacted, scalarized boson stars exist in our model.

\section{A tachyonic instability for boson stars}
\label{background}
The spontaneous scalarization of a boson star in a model containing
a non-minimal coupling between the scalar field and the Ricci scalar ${\cal R}$ and/or the Gauss-Bonnet term ${\cal G}$, respectively, can be related to the appearance of a tachyonic instability of the scalar field $\phi$. 
We assume the fixed, background space-time to be that of a boson star, which is a solution
to the coupled, non-linear set of equations (\ref{eom}) for $\alpha=\gamma=0$. The explicit form of these equations reads~:
\begin{eqnarray}
\label{BS_background}
N'&=&-r\left(m^2 \psi^2 + N\psi'^2 + \frac{\omega^2\psi^2}{
     \sigma^2 N}\right)-\frac{N}{r}+\frac{1}{r}  \ \ \ , \ \ \
\sigma'= r \left(\sigma\psi'^2+\frac{\omega^2\psi^2}{N^2\sigma}\right) \nonumber \\
\psi''&=& \frac{m^2\psi}{N}\left(r \psi \psi'+ 1\right) - \frac{\omega^2\psi}{N^2\sigma^2}
     -\frac{\psi'}{r}\left(1+\frac{1}{N}\right)  \ ,
\end{eqnarray}
where now and in the following the prime denotes the derivative with respect to $r$.

The boundary conditions for the system are determined by the requirement of finite energy and global regularity of the functions. These read
\begin{equation}
\label{bc_origin}
      N(0) = 1 \ \ , \ \  \left.\frac{d\psi}{dr}\right\vert_{r=0} = 0  
\end{equation}
at the origin $r=0$ and 
\begin{equation}    
\label{bc_infinity_all}
\sigma(r \to \infty) \rightarrow 1 \ \ , \ \ \psi(r \to \infty) \sim \frac{1}{r}\exp\left(-  \sqrt{m^2 - \omega^2}r \right) \rightarrow 0 \ 
\end{equation}
at spatial infinity.

In order to study the spontaneous scalarization of boson stars, let us first recall the pattern of solutions to the equations (\ref{BS_background}). For that, note that the asymptotic behaviour of the scalar
function $\psi(r)$ restricts the existence of solutions to $\omega/m \in [\omega_{\rm min}/m,1.0]$. The value of
$\omega_{\rm min}/m$ has to be determined numerically and is $\omega_{\rm min}/m\approx 0.76$. 
For $\psi(0)\equiv \psi_0=0$, the branch of boson star solutions is connected to the Minkowski vacuum
where $N(r)\equiv 1$, $\sigma\equiv 1$, $\psi\equiv 0$ and $\omega/m=1.0$. Increasing $\psi_0$ 
decreases the value of $\omega/m$  down to $\omega_{\rm min}/m$ from where a second branch of
solutions exists on which $\psi_0$ increases further and $\omega/m$ increases again up to $\omega_{\rm cr,1}/m\approx 0.86$. From there, a succession of branches exists with decreasing extent of the $\omega$-interval that form the characteristic spiral for boson star solutions. At the center of this spiral, $\sigma(0)\rightarrow 0$,
while $\psi_0\rightarrow \infty$, i.e. the spiral ends in a singular solution. 
The stability of boson stars with respect to decay into $Q$ individual bosons can be investigated by comparing
the mass $M$ of the solution with the mass of $Q$ individual bosons with mass $m$.

In the following, we will assume the background space-time of the boson star to be fixed. We study the
equation resulting from the variation of the action (\ref{action}) with respect to the real scalar field $\phi$.
In the following, we will assume $\phi=\phi(r)$ corresponding to the lowest angular mode in $\theta$, i.e. $\ell=0$.
The equation for $\phi$ then takes the form~:
\begin{equation}
\label{eq:klein_gordon}
    \frac{1}{\sigma r^2} \left(r^2\sigma N \phi'\right)' = -\left(\alpha \cal{R} + \gamma {\cal G}\right)\phi   \ \ \  ,
\end{equation}
where the Ricci scalar ${\cal R}$ and Gauss-Bonnet term ${\cal G}$ for the metric (\ref{eq:metric}) read, respectively~:
\begin{eqnarray}
{\cal R}= -\frac{2N}{r^2} - \frac{4N'}{r} - N'' + \frac{2}{r^2} - \frac{4\sigma'}{\sigma} \frac{N}{r} - \frac{3 \sigma'}{\sigma} N' - \frac{2\sigma''}{\sigma}N \ ,
\end{eqnarray}
and
\begin{eqnarray}
{\cal G}= \frac{4}{r^2}\left(N N'' + N'^2 - N'' + \frac{5\sigma'}{\sigma}N N'
- \frac{3 \sigma'}{\sigma}N'+ \frac{2\sigma''}{\sigma}N^2 - \frac{2\sigma''}{\sigma}N\right) \ .
\end{eqnarray}

Since the background space-time is fixed and a solution to the Einstein equation $G_{\mu\nu}=T_{\mu\nu}^{(\psi)}$, we can use the
identity ${\cal R}=-{\cal  T}^{(\psi)}$ to replace the Ricci scalar ${\cal R}$ in terms of the trace of 
the energy momentum tensor of the scalar field $\psi$, ${\cal T}^{(\psi)}$~:
\begin{equation}
    -{\cal R} =  {\cal T}^{(\psi)} = \left(\frac{2 \omega^2}{N \sigma^2} - 4 m^2 \right) \psi^2 - 2 N (\psi')^2 \ .
\end{equation} 
The equation (\ref{eq:klein_gordon}) can be solved with the boundary conditions for the field $\phi$, which are
\begin{equation}
\left.\frac{d\phi}{dr}\right\vert_{r=0} = 0 \ \ \ , \ \ \ \phi(r\rightarrow\infty)\rightarrow 0 \ .
\end{equation}
Note also that the equation is linear in $\phi$, so we can choose $\phi(0)=1$ without loss of generality.

For completeness let us remark that (\ref{eq:klein_gordon}) can be put into the form of
a Schr\"odinger-like equation~: 
\begin{equation}
\label{eq:schrodinger}
      \frac{d^2\chi}{dr_*^2} = V_{\rm eff} \chi\ \  {\rm with} \ \ V_{\rm eff} = \sigma^2 N\left(\frac{\sigma'}{\sigma} \frac{N}{r} + \frac{N'}{r} + \alpha {\cal T}^{(\psi)}-\gamma {\cal G}\right)
\end{equation}
where $r_*$ is related to $r$ via ${\rm d}r/{\rm d}r_*=\sigma N$ and
$\chi(r):=r\phi(r)$.  The tachyonic instability corresponds to the
existence of bound states of the Schr\"odinger-like equation. In the following, we will demonstrate that
these bound states do exist. Obviously, we would not expect a non-trivial solution for $\phi$ for generic values of $\alpha$
and $\gamma$, respectively.
We have solved
the equation (\ref{eq:klein_gordon}) numerically in the background of the (numerically given) boson star solutions and find that boson stars can be scalarized for specific choices of $\alpha$ and $\gamma$.

\begin{figure}[ht!]
\begin{center}
\input{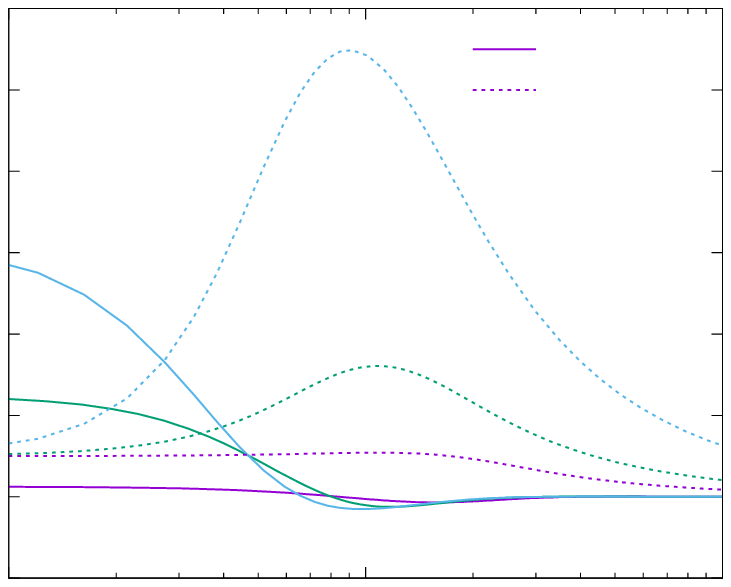}
\input{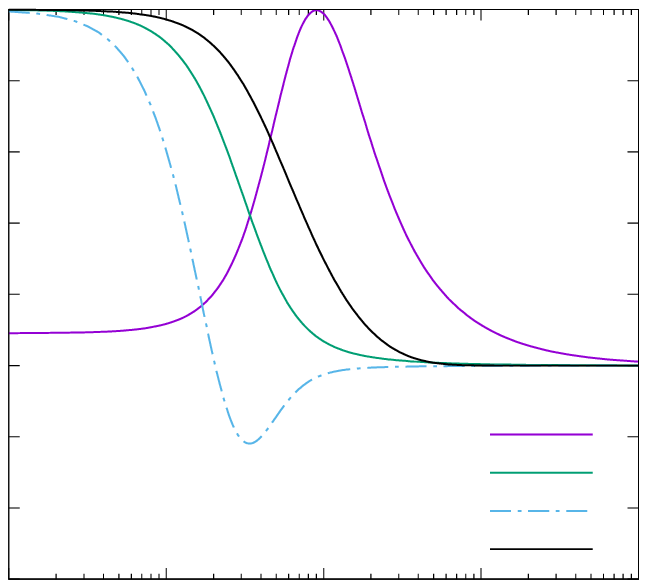}
\caption{We show the effective potential $V_{\rm eff}$ (solid) of (\ref{eq:schrodinger}) for three different values of  $\psi_0$ (see Table \ref{fig1} for the corresponding values of $\alpha$, $\sigma(0)$ and $\omega$)
and $\alpha > 0$, $\gamma=0$ (left). We also show the profile of the scalar field function $\phi(r)$ in a boson star background with  $\psi_0 = 1.0$,
which corresponds to $\omega=0.812$. Two solutions without nodes (i.e. $k=0$) of (\ref{eq:klein_gordon}) exist in this case : one for $\alpha = 5.160 $ (purple solid)
and one for $\alpha = -1.840$ (green solid). We also show the profile for $\phi(r)$ for the first excited
solution, $k=1$, which corresponds to $\alpha=-8.788$ (blue dotted-dashed). For clarification, we also plot the profile of the boson star scalar
field $\psi(r)$ (solid black). Note that $\phi(r)$ has been divided by its maximal value for the $k=0$, $\alpha=5.160$ case. }
\label{fig1}
\end{center}
\end{figure}

\subsection{$\gamma=0$}
This corresponds to the scalarization of the boson star due to the $\alpha\phi^2 {\cal R}$ term.
This model has been studied previously in \cite{ref2new,ref3new} and it was demonstrated that
boson stars can spontaneously develop scalar hair for appropriate choices of $\alpha > 0$. In this paper, we confirm the obtained results for $\alpha > 0$, but demonstrate that boson stars can be scalarized also for $\alpha < 0$. Moreover, we find that radially excited scalar hair exists. In the following, we will denote the number of nodes in the scalar field
function $\phi(r)$ by $k$ with $k=0$ being the fundamental mode.

\begin{figure}[ht!]
\begin{center}
\input{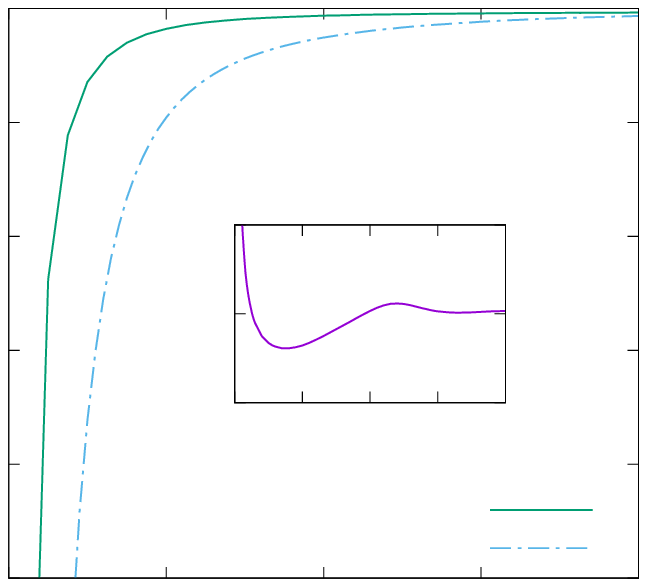}
\input{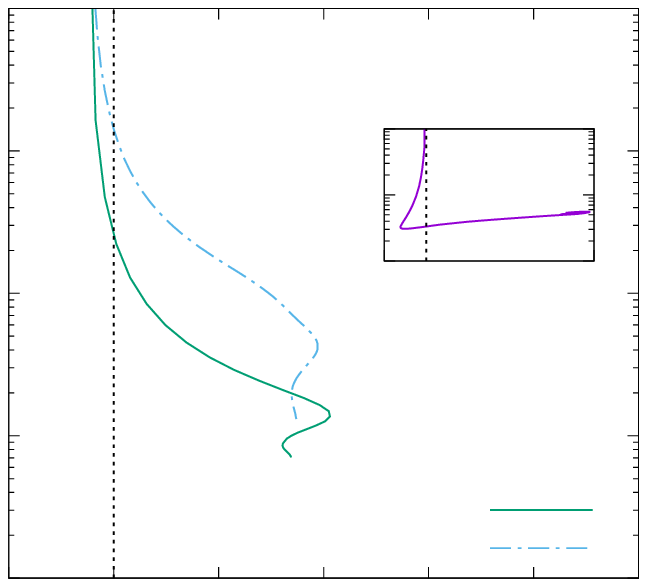}
\caption{We show the dependence of the value of the parameter $\alpha$ for which non-trivial solutions of (\ref{eq:schrodinger}) exist, on the central value of the complex scalar field forming the boson star, $\psi_0$ (left), as well as on the ratio between the mass $M$ and the Noether charge $Q$ (right). We present our results for the solutions without nodes, $k=0$, for positive $\alpha$ (solid purple, $+$, see insets of both figures) as well as for negative 
$\alpha$ (solid green, $-$) and for the first excited solution, $k=1$, (blue dotted-dashed), respectively. For clarity, we
indicate the $M/Q=1$ line (dotted black, right).}
\label{fig2}
\end{center}
\end{figure}

\begin{table}
\begin{center}
\begin{tabular}{|c|c|c|c|c|c|c|}
\hline
$\psi_0$ & $\omega$ & $\sigma(0)$ & ($\alpha$, $\gamma$) & ($\alpha$, $\gamma$) & ($\alpha$, $\gamma$) & ($\alpha$, $\gamma$)\\
\hline
$1.0$ & $0.812$ & $0.190$ & ($5.160$, $0$) & ($-1.841$, $0$) &  ($0$, $0.782$) & ($0$, $-0.053$) \\
$0.8$ & $0.774$ & $0.287$ & ($4.335$, $0$) & ($-3.555$, $0$) &  ($0$, $2.338$) & ($0$, $-0.219$)  \\
$0.5$ & $0.785$ & $0.484$ & ($3.217$, $0$) &  ($-47.554$, $0$) &   ($0$, $14.777$) & ($0$, $-1.926$)   \\
\hline
\end{tabular}
\caption{Values of  $\psi_0$, frequency $\omega$ and $\sigma(0)$, i.e. the value of the metric function $\sigma(r)$ at the origin, of
the background boson star solution and values of ($\alpha$, $\gamma=0$) and ($\alpha=0$, $\gamma$), respectively, for which non-trivial solutions without nodes ($k=0$) of (\ref{eq:klein_gordon}) exist.}
\label{table1}
\end{center}
\end{table}

In Fig. \ref{fig1} (left), we show the effective potential $V_{\rm eff}(r)$ as given in (\ref{eq:schrodinger}) and the corresponding bound state
solution for positive values of $\alpha$ and $k=0$. The value of  $\psi_0$, 
the frequency $\omega$ and $\sigma(0)$, the value of the metric function
$\sigma(r)$ at $r = 0$, of the background boson star as well as the value of $\alpha$, for which bound states exist, are given in Table \ref{table1}. The value of the effective potential $V_{\rm eff}(r = 0)$ increases with increasing  $\psi_0$ or –
equivalently – decreasing $\sigma(0)$. Moreover, $V_{\rm eff}$ develops a local, negative valued minimum with increasing
$\psi_0$. This leads to a developed local maximum of the scalar field $\phi(r)$ at a finite value of $r$, which corresponds
roughly to the outer radius of the boson star. To state it differently~: the boson star gets scalarized strongest on a shell
around the outside core of the star. 
In the same Fig. \ref{fig1} (right) we show the example of a scalarization for a boson star with $\psi_0=1.0$ which
corresponds to $\omega=0.812$. While scalarization of boson stars for positive $\alpha$ (denoted
by $k=0, +$ in Fig. \ref{fig1} (right)) has been
demonstrated before, we present here the first results for a scalarization of boson stars with negative $\alpha$
 (denoted by $k=0, -$ in Fig. \ref{fig1} (right)).  The existence of these solutions is related to the fact that
the effective potential $V_{\rm eff}$  becomes negative close to the origin of the coordinate system  for sufficiently large values of $\psi_0$ and hence a tachyonic instability exists
for $\alpha < 0$. To give an idea on the difference between the solutions, we show the profile of $\phi(r)$ for a
boson star background with  $\psi_0 = 1.0$ for the two possible solutions with $k=0$
in Fig. \ref{fig1} (right).  Obviously, the scalar field profiles $\phi(r)$ are very different in the two cases~: in contrast to the case discussed above, the $\alpha < 0$ scalarization happens completely {\it inside}
 the boson star with the maximum of the scalar field $\phi(r)$ at the center $r=0$.  
 In addition, the profile of $\phi(r)$ for $k=1$ suggests that this radially excited solution is rather an excitation
 of the $\alpha < 0$ solution than of the $\alpha > 0$ solution. Correspondingly, the excited solution exists only for negative $\alpha$. 

The qualitative features of the scalarization for $\alpha > 0$ as compared to $\alpha < 0$ can further
be understood when considering the value of $\alpha$ for which scalarization is possible in dependence
on the central value of the boson star, $\psi(r=0)\equiv \psi_0$ and the ratio between the mass $M$ and the Noether charge $Q$, respectively. This is shown in Fig. \ref{fig2}.
For $\alpha < 0$, the boson star can be scalarized for $\psi_0 \gtrsim 0.4$ and $\alpha$ is a monotonically
increasing function of $\alpha$.  For $\alpha > 0$, boson stars with arbitrarily small $\psi_0$ can
be scalarized with $\alpha\rightarrow +\infty$ for $\psi_0\rightarrow 0$, however, $\alpha$ is not
a monotonic function of $\psi_0$ and in particular possesses a global
minimum for finite central density indicating a {\it gap}, $\alpha\in [0:3]$, for which boson stars can {\it never} be scalarized, independent of their mass or central density.  The dependence of $\alpha$ on $\psi_0$ for the $k=1$ solution is qualitatively similar to that of the $\alpha < 0$ branch, 
Fig. \ref{fig2} indicates that
$\psi_0 \gtrsim 0.4$ for the excited solution to exist.  Moreover, we find that the $k=0$, $\alpha < 0$ and the
$k=1$ solutions  appear approximately
in the interval $[\omega_{\rm min} : \omega_{{\rm cr},1}]$, i.e. the interval of existence of the second branch of boson star solutions. In Fig.\ref{fig2} (right) we show $\alpha$ in dependence on the ratio $M/Q$. This ratio indicates
whether the background boson star solution is stable or unstable to decay into $Q$ bosons of mass $m\equiv 1$.
We find that boson stars that are stable with respect to such a decay, i.e. solutions with $M/Q < 1$, can be scalarized
for positive {\it and} negative values of $\alpha$. 

Finally, Table \ref{table1} contains the values of $\alpha$ for some examples of boson stars. The numbers clearly indicate
that scalarization requires increasing $\vert\alpha\vert$ when decreasing $\psi_0$ for scalarization to appear.
Or in other words~: low mass boson stars are harder to scalarize.

\subsection{$\alpha=0$ case}
This case has not been studied previously. The spontaneous scalarization of neutron stars (NS) has been discussed briefly in \cite{ref22} and it was found that NS can be scalarized for both signs of $\gamma$. This is also true for boson stars as we will demonstrate below. 

\begin{figure}[ht!]
\begin{center}
\input{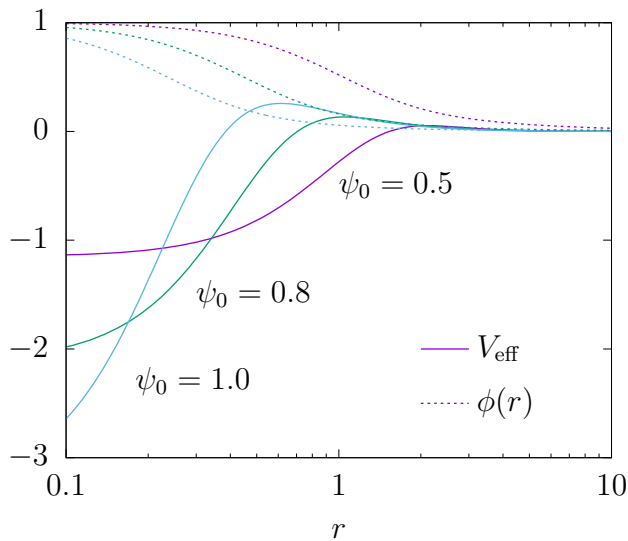}
\caption{We show the effective potential $V_{\rm eff}$ (solid) of (\ref{eq:schrodinger}) for three different values of  $\psi_0$ for $\gamma < 0$, $\alpha=0$. 
We also give the corresponding bound state solution $\phi(r)$
for these same values of  $\psi_0$ (dashed). For values of $\alpha$ and $\gamma < 0$, respectively, see Table \ref{table1}. }
\label{fig1b}
\end{center}
\end{figure}

The qualitative features of the solutions
for $\gamma \gtrless 0$, $\alpha=0$ are equivalent to the ones with $\alpha \gtrless 0$, $\gamma=0$. 
In Fig. \ref{fig1b} we show the effective potential $V_{\rm eff}$ of (\ref{eq:schrodinger}) for $\gamma < 0$.
The solutions $\phi(r)$ are maximal at the center of the boson star and the scalarization happens pratically completely inside the star. Correspondingly, the branch for positive $\gamma$ shows functions of $\phi(r)$ with a local
maximum close to the outside core of the star. Hence, in this sense, scalarization seems similar to the
case $\alpha\neq 0$, $\gamma=0$.

\begin{figure}[ht!]
\begin{center}
\input{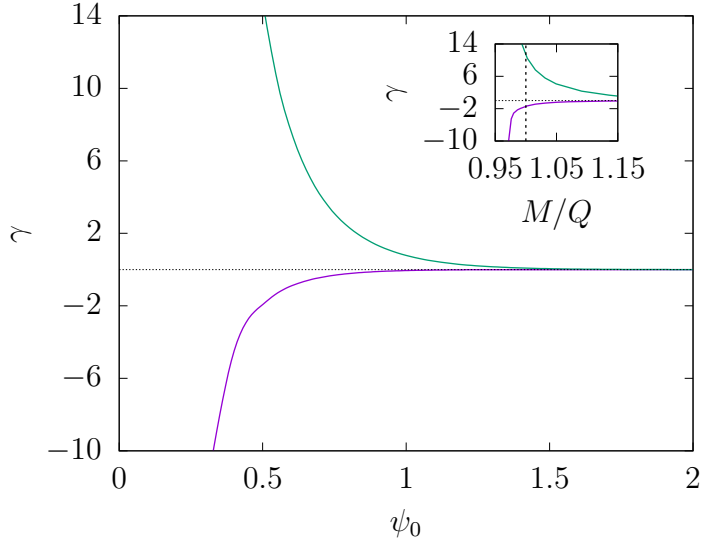}
\caption{We show the value of $\gamma$ in dependence on the central boson star value $\psi_0$ for which
solutions of (\ref{eq:klein_gordon}) exist when $\alpha=0$.  In the inset, we also show $\gamma$ in dependence on the
ratio between the mass $M$ and the Noether charge $Q$ of the background boson star. Note that boson stars with $M/Q < 1$ are stable with respect to decay into $Q$ bosons of mass $m\equiv 1$. For clarity, we indicate the $M/Q=1$ line (black dotted).  }
\label{fig:gamma_alpha0}
\end{center}
\end{figure}

However, we find that an important difference between the $\alpha$- and $\gamma$-scalarization exists.
While for $\gamma=0$, boson stars can never be scalarized for $\alpha\in [0:3]$, a similar gap in $\gamma$ is {\it not} present for $\alpha=0$. This means that -- in principle and choosing the central density of the boson star sufficiently large (see below) -- boson stars can be scalarized for all values of $\gamma\neq 0$. 
This is shown in Fig. \ref{fig:gamma_alpha0}, where we give $\gamma$ in dependence of $\psi_0$ as well as
$\gamma$ in dependence on the ratio $M/Q$ (inset).  For some specific values of $\psi_0$ see also
Table \ref{table1}.  Again, boson stars that are {\it a priori} stable with respect to the decay into $Q$ bosonic
particles, i.e. those solutions with $M/Q < 1$ can be scalarized for both signs of $\gamma$ (see inset of Fig. \ref{fig2}).
Again, scalarization is harder for low central values of $\psi$, but in contrast to the $\alpha$-scalarization, $\gamma$
can be chosen arbitrarily small (positive or negative) for sufficiently large $\psi_0$.

\begin{figure}[ht!]
\begin{center}
\input{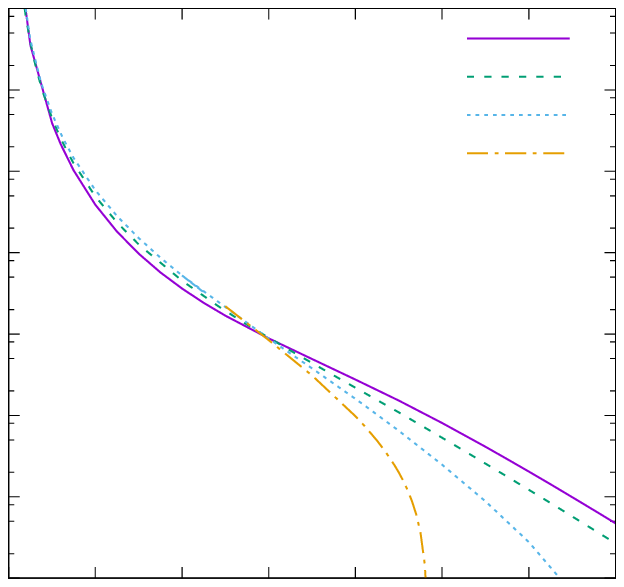}
\input{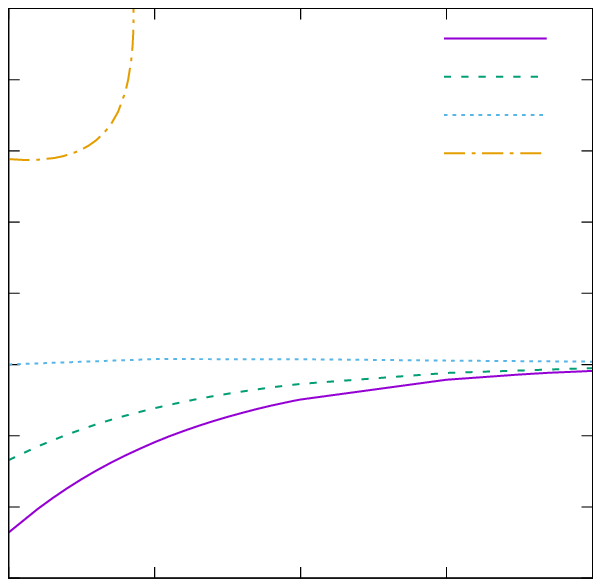}
\caption{
We show the dependence of the value of the parameter $\gamma < 0$ for which non-trivial solutions of (\ref{eq:klein_gordon})
exist, on the central value of the complex scalar field forming the boson star, $\psi_0$, for different values of $\alpha$. The left figure shows the interval $\psi_0\in [0:1.4]$ , while
the right figure shows the interval $\psi_0\in [1.4:1.8]$.}
\label{fig_new1}
\end{center}
\end{figure}

\subsection{$\alpha\neq 0$, $\gamma\neq 0$}
We have also studied the influence of the two terms that can -- separately -- lead to scalarization
of the boson star onto each other. For that we have chosen a boson star with $\psi(0)\equiv\psi_0=1.0$.
 In this case, as discussed above, two solutions without nodes in $\phi(r)$ exist,
one for positive value of $\alpha$ ($\gamma$) and one for negative value of $\alpha$ ($\gamma$), respectively. 
We have first studied the $\gamma$-scalarization for fixed values of $\alpha$. Our results are shown in Fig.\ref{fig_new1},
where we give the value of $\gamma$ in dependence on $\psi_0$ for four different values of $\alpha$.
As the curve for $\alpha=0$ indicated, we have deformed the solutions of the negative $\gamma$ branch.
In the interval $\psi_0\in [0:1.4]$ the qualitative features do not change much when choosing $\alpha=-2,-1,0,1$, respectively~:
$\gamma\rightarrow -\infty$ for $\psi_0\rightarrow 0$, while $\gamma\rightarrow +\infty$ for $\psi_0\rightarrow +\infty$ 
and $\alpha=-1,0,1$.  However, for $\alpha$ sufficiently negative, in our example $\alpha=-2$, a critical value of $\psi_0=\psi_{0,{\rm cr}}(\alpha)$ exists,
at which $\gamma=0$. We find that $\psi_{0,{\rm cr}}(\alpha=-2)\approx 0.97$. Increasing $\psi_0$ from this value requires
$\gamma > 0$ for scalarization to appear. However, $\psi_0$ can only be increased up to $\psi_0\approx 1.49$, where $\gamma\rightarrow +\infty$.

\begin{figure}[ht!]
\begin{center}
\input{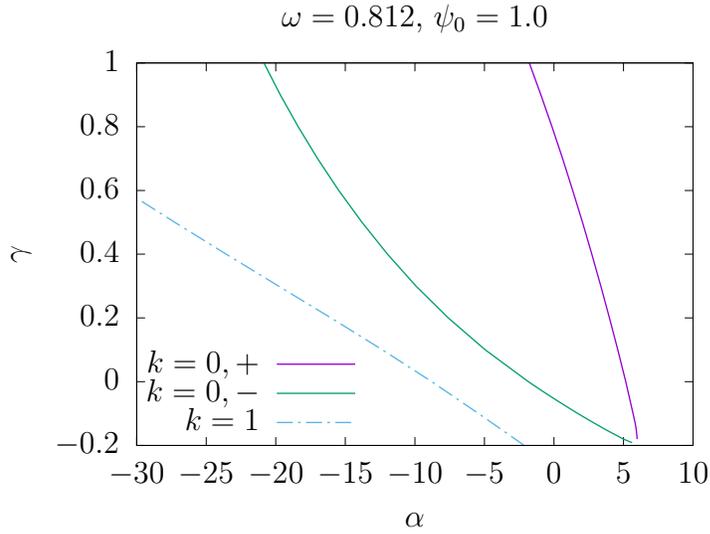}
\caption{We show the value of $\gamma$ in dependence on $\alpha$ for which solutions
to (\ref{eq:klein_gordon}) exist in the background space-time of a boson star with $\psi_0=1.0$ corresponding to $\omega=0.812$. We show the two branches that exist for $k=0$ ($+$: purple solid and
$-$: green solid, respectively) as well as the branch of solutions with one node, i.e. $k=1$ (blue dotted-dashed).
Note that $+$ and $-$ relate to the solutions that exist for positive and negative $\alpha$, respectively, in
the $\gamma=0$ limit.} 
\label{fig_new2}
\end{center}
\end{figure} 

To get a further understanding of the interplay between the two terms, we have chosen the two solutions without nodes, $k=0$, as well as the $k=1$ solution, respectively, that exist for $\gamma=0$ and studied their evolution in function of $\gamma$. This
is shown in Fig. \ref{fig_new2} for a boson star solution with $\psi_0=1.0$. Decreasing
$\gamma$ clearly shifts the values of $\alpha$ necessary for scalarization on both branches for $k=0$ as well as for the $k=1$ solution to larger values. Moreover, the difference between the value of $\alpha$ needed
for scalarization on the $k=0, +$ branch and the $\alpha$ needed on the $k=0,-$ branch decreases with decreasing 
$\gamma$. For sufficiently small $\gamma=\gamma_{\rm eq}$, the $+$ branch and the $-$ branch join. For $\psi_0=1.0$, we find that 
$\gamma_{\rm eq}\approx -0.2$ with $\alpha_{\rm eq}\approx 6$.

%%%%%%%%%%%%%%%%%%%%%%%%%%%%%%%%%%%%%%%%%%%%%%%%%

\section{Influence of backreaction}

\begin{figure}[ht!]
\begin{center}
\input{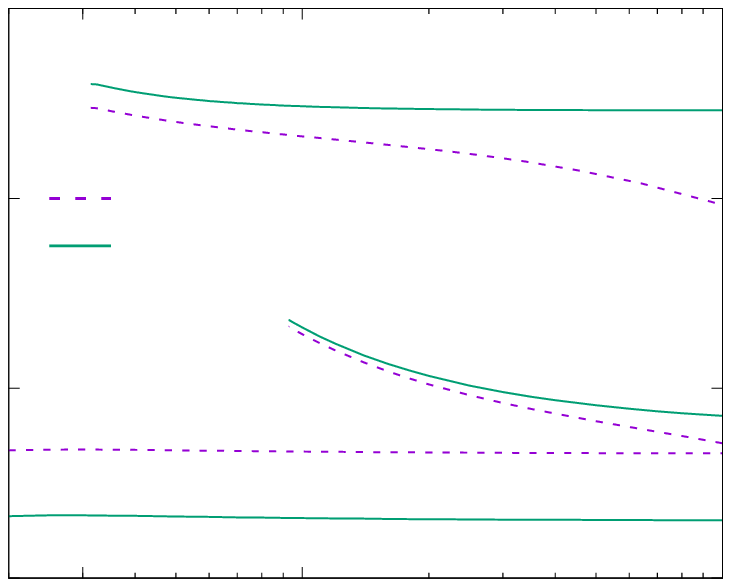}
\input{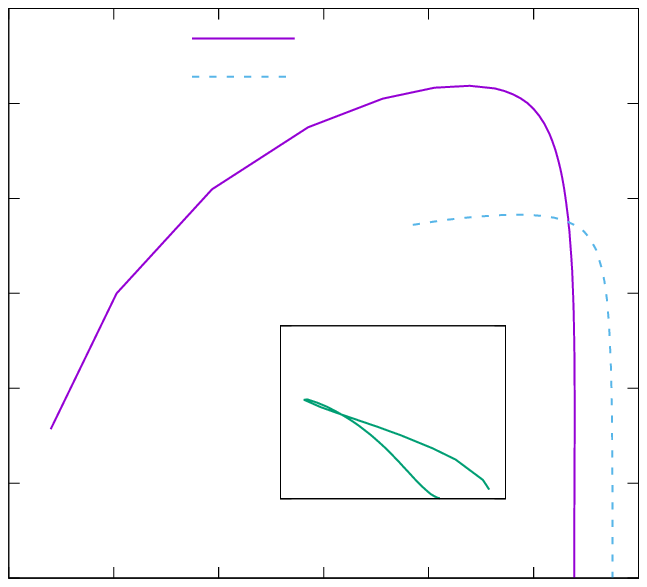}
\caption{Left: We show the mass $M$ (purple dashed) and the Noether charge $Q$ (green solid) of the fully
backreacted scalarized boson star as function of the scalar-tensor coupling $\vert\alpha\vert$ for two different values of
the frequency $\omega$ (right) including the positive and negative $\alpha$ branch, respectively, for $\omega=0.85$. Note that
for $\omega=0.96$ only one branch of solutions exists. 
Right: We show the scalar charge $Q_{\phi}$ in function of $M/Q$ for $\omega=0.85$ (solid) and
$\omega=0.96$ (dashed). For $\omega=0.85$ we give the values for the positive $\alpha$ branch (solid purple)
and for the negative $\alpha$ branch (solid green, inset), respectively.}
\label{fig4}
\end{center}
\end{figure}

The results obtained in the previous section indicate that scalarization of boson stars is possible. In the
following, we demonstrate that, indeed, scalarized boson stars exist when taking the full backreaction of
the space-time into account.  We have solved the equations (\ref{eom}) numerically with the appropriate boundary conditions for the case $\gamma=0$. 

\begin{figure}[ht!]
\begin{center}
\input{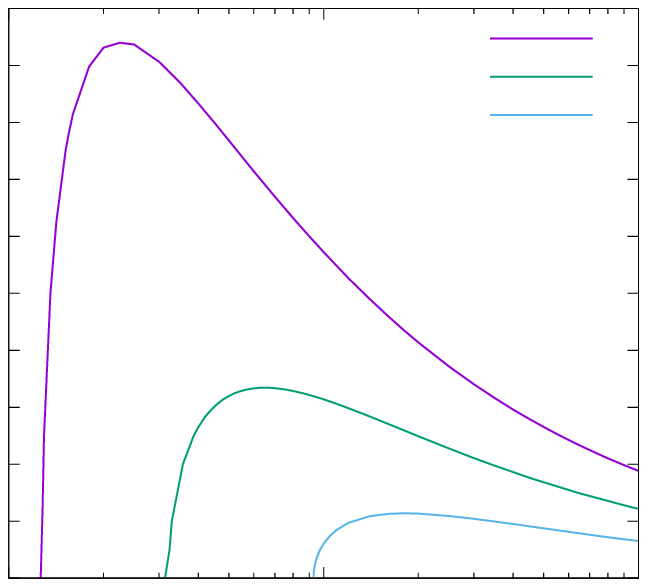}
\input{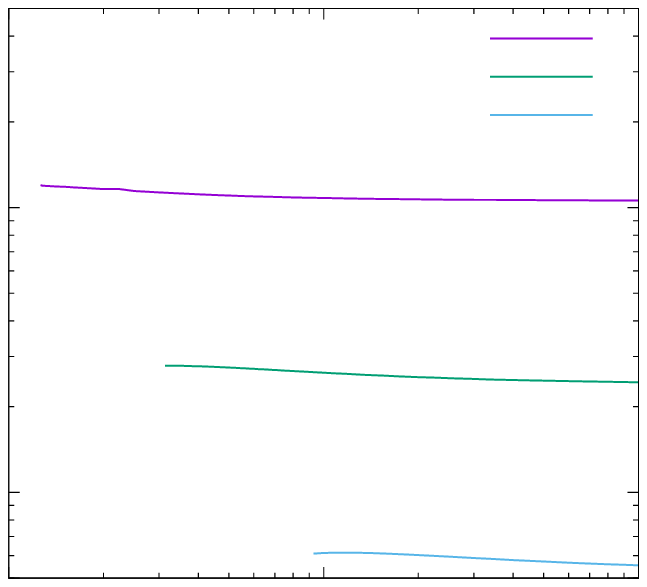}
\caption{We give the value of  $\phi_0$ (left) and $\psi_0$ (right) as function of $\vert\alpha\vert$  for the positive (purple solid) and negative (green solid) $\alpha$ branch, respectively, for $\omega=0.85$. For comparison, we also
give the values for $\omega=0.96$.   }
\label{fig5}
\end{center}
\end{figure}

To study the effect, we have fixed the value of the frequency $\omega$ and studied
the dependence of the parameters describing the solution in function of $\alpha$. We have concentrated on the main branch of solutions, choosing $\omega=0.96$ and $\omega=0.85$, respectively, which corresponds to $\psi_0\approx 0.05$ and $\psi_0\approx 0.28$ 
with $\phi_0$ small, for which the non-backreacted case is a good approximation. Using these
solutions we have then increased the value of $\phi_0$ to study the influence of backreaction.
Our results are show in Fig. \ref{fig4} and Fig. \ref{fig5}. For $\omega=0.85$, the scalarized boson star exists for positive and negative
$\alpha$, while for $\omega=0.96$ only solutions with $\alpha > 0$ are possible. 
Here, we give the mass $M$ and the Noether charge $Q$ for these solutions in function of $\vert\alpha\vert$.
As in the background case, we find that a {\it gap} exists in $\alpha$ for which scalarized boson stars
do not exist. This gap decreases with decreasing $\omega$. This can also be seen in Fig.\ref{fig5}, where we give
the value of $\phi(r)$ at the origin, $\phi_0$ (left), as well as the boson star field $\psi(r)$ at the origin, $\psi_0$ (right),
as function of $\vert\alpha\vert$. While $\psi_0$ varies little when increasing $\vert\alpha\vert$, we find that $\phi_0\rightarrow 0$ at a critical value of $\vert\alpha\vert_{\rm cr}$. For $\omega=0.85$, we find
$\alpha_{\rm cr}\approx 3.2$ and $\alpha_{\rm cr}\approx -1.3$, while for $\omega=0.96$ our results indicate
that $\alpha_{\rm cr}\approx 9.3$. Hence, for $\omega=0.85$ a {\it gap} in $\alpha$ exists, $\alpha\in [-1.3:3.2]$ for which
scalarized boson stars do not exist. For $\omega=0.96$, we need to require $\alpha \gtrsim 9.3$.  

Fig. \ref{fig4} also demonstrates that for solutions with positive $\alpha$, the ratio $M/Q < 1$ always, indicating that the presence of the extra gravity scalar starts to increase the binding between the individual 
bosons in the star. On the other hand, the solutions of the negative $\alpha$ branch are unstable to decay into
$Q$ individual bosons. The behaviour of the scalar charge $Q_{\phi}$ 
associated to the fall-off of the gravity scalar $\phi$ at infinity, see (\ref{bc_infinity_all}), is 
qualitatively very different when comparing the $\alpha > 0$ and the $\alpha < 0$ solutions. 
Fig.\ref{fig4} (right) demonstrates that for $\alpha > 0$
and for a fixed value of $M/Q$, the scalar charge $Q_{\phi}$ characterises the scalarized boson star uniquely. 
Increasing $M/Q$, the scalar charge increases up to a maximal value and then decreases sharply to zero
at the maximal possible value of $M/Q$ for these solutions. Moreover, the smaller $\omega$, the larger the maximal
possible scalar charge $Q_{\phi}$. 
The solutions with $\alpha <0$ show a very different behaviour for $Q_{\phi}$, see inset of Fig.\ref{fig4} (right).
The scalar charge is much smaller as compared to the $\alpha > 0$ branch and no longer uniquely characterizes the solution
since for a small interval in $M/Q$ two solutions with different $Q_{\phi}$ exist when fixing $M/Q$.

\section{Conclusions}
Spontaneous scalarization of black holes in scalar-tensor gravity models has been discussed extensively 
recently. Since static, electro-vacuum, asymptotically flat black holes fulfil ${\cal R}\equiv 0$,
these black holes cannot be scalarized in a $\alpha\phi^2{\cal R}$ model, while scalarization in a $\gamma\phi^2{\cal G}$ model is
possible. 

In this paper, we have studied ultra-compact objects that are globally regular in a $(\alpha{\cal R} + \gamma{\cal G})\phi^2$
model and demonstrated that they can be scalarized for both signs of $\alpha$ and $\gamma$, respectively. 
While there exists a gap in $\alpha$ for which boson stars can never be scalarized when $\gamma=0$, this is not present in
the converse case. 
The nature of the scalarization for positive as compared to negative values of the couplings is very different~:
while for positive couplings the scalarization happens mainly on a shell around the outer core of the boson
star, it is mainly the interior of the boson star that gets scalarized for negative couplings. 
Moreover, the scalar field in the negative coupling case can be excited. It would be interesting to understand what
such an excited boson star would decay into and what type of radiation it would emit. This radiation, if detected,
could be a distinguishing feature of compact objects as compared to black holes. 

When taking the backreaction for the $\alpha$-scalarization into account, the ratio between mass $M$ and Noether charge $Q$
indicates that scalarized boson stars with $\alpha >0$ are stable to decay into $Q$ individual bosons, while
those for negative $\alpha$ are unstable with respect to this decay. Moreover, the $\alpha > 0$ boson stars
can be uniquely characterized by the ratio $M/Q$ as well as the scalar charge $Q_{\phi}$, which is defined
by the long-range fall-off of the gravity scalar at infinity. It will be interesting
to see whether such a unique description is possible in the case of gauging the U(1) symmetry of the model, i.e.
when electrically (and/or magnetically) charging the boson stars. This is currently under investigation. 

\vspace{1cm}

{\bf Acknowledgements} 
 BH would like to thank FAPESP for financial support under
grant numbers {\it 2016/12605-2} and {\it 2019/01511-5}, respectively, as well as CNPq for financial support under {\it Bolsa de Produtividade Grant 304100/2015-3}.  BH would also like to thank the research group
{\it Physique th\'eorique et math\'ematique} at the Universit\'e de Mons (Belgium) for their hospitality.

\vspace{2cm}

%%%%%%%%%%%%%%%%%%%%%%%%%%%%%%%%%%%%%%%%%%%%%%%%%%%%%%%%%%%%

 \end{document}